# Making it possible: constructing a reliable mechanism from a finite trajectory

Ophir Flomenbom

*Flomenbom-BPS, 19 Louis Marshal, Tel-Aviv, Israel 62668*

## ABSTRACT

Deducing an underlying multi-substate *on-off* kinetic scheme (KS) from the statistical properties of a two-state trajectory is the aim of many experiments in biophysics and chemistry, such as, ion-channel recordings, enzymatic activity and structural dynamics of bio-molecules. Doing so is almost always impossible, as the mapping of a KS into a two-state trajectory leads to the loss of information about the KS (almost always). Here, we present the optimal way to solve this problem. It is based on unique forms of reduced dimensions (RD). RD forms are *on-off* networks with connections only between substates of different states, where the connections can have multi-exponential waiting time probability density functions (WT-PDFs). A RD form has the simplest topology that can reproduce a given data. In theory, only a single RD form can be constructed from the full data (hence its uniqueness), still this task is not easy when dealing with finite data. For doing so, a toolbox made of known statistical methods in data analysis and new statistical methods and numerical algorithms developed for this problem is presented. Our toolbox is self-contained: it builds a mechanism based *only* on the information it extracts from the data. The implementation of the toolbox on the data is fast. The toolbox is automated and is available for academic research upon electronic request.

## KEYWORDS

**Two-state trajectories**: analysis of-, mechanisms from-, canonical forms of-
**Single molecules**: trajectories of-, time traces of-, analysis of-
**Theory of Biophysical processes**: ion channels, enzymes, DNA & RNA dynamics, diffusion, blinking of nano-crystals

# CONTENTS

# I. Introduction

High resolution experiments, in which one measures the dynamics of individual entities during activity, give, in many cases, time trajectories made of several states. The simplest example of such trajectories is an *on-off* trajectory made of *on* and *off* periods (waiting times); see Fig. 1A. Examples of processes that give (in some circumstances) such data include the passage of ions and biopolymers through individual channels [3-5], activity and conformational changes of biopolymers [1-2, 6-16], diffusion of molecules [17-20], and blinking of nano-crystals [21-24]. Given an *on-off* trajectory, one tries to deduce the mechanism of the observed process. (In experiments, the data is noisy; here, we assume that a noiseless trajectory can be obtained.) Usually, the fundamental conjecture is that the mechanism of the observed process can be described by a multi-substate *on-off* Markovian kinetic scheme (KS) [25-39]; see Fig. 1B. (In Refs. [40-56] related descriptions for relevant processes are discussed.) The *on-off* KS can stands for one of the following physical systems: a discrete conformational energy landscape of a biomolecule, steps of a chemical reaction with conformational changes or environmental changes, a set quantum states, etc.

Within these assumptions, the measured stochastic signal represents a random walk in the *on-off* KS, where only transitions between substates of different states in the *on-off* KS leads to a change in the value of the stochastic signal.

From single molecule experiments, we wish to learn as much as possible about the underlying KS. Clearly, we wish to learn much more about the KS when performing and analyzing such complicated experiments than the general properties deduced about the KS from bulk measurements. This task is not so hard: it is simple to show that bulk measurements are equivalent to only one property of the *on-off* trajectory[1], whereas the *on-off* trajectory contains many other unique properties. Therefore, the following questions arise: how to extract all the information from the data? (and, here, it is important to know the maximal information content in the data), and, how to translate the extracted information into a reliable mechanism that can generate the data?

To better define the problem, we straightforwardly assume that the information content in the data is known, and focus on the problem of translating the information content in the data into a mechanism. If we try to build a KS from the data, we find that determining the KS from the two-state trajectory is difficult. There are several reasons for this: first, the number of the substates in each of the states, $L_x$ ($x = on$, *off*), is usually large, and the connectivity among the substates is usually complex. The data however has limited information content, and so *all* the details of a KS can't be determined from the limited data. A fundamental difficulty in finding the correct KS arises from the equivalence of KSs; namely, there are a number of KSs that lead to the same trajectories in a statistical sense [28-30, 32-36].

A way to deal with these issues is to use canonical (unique) forms rather than KSs [32-35]: the space of KSs is mapped onto a space of canonical forms. A given KS is mapped into a unique canonical form but many KSs can be mapped to the same canonical form. Underlying KSs with the same canonical form are equivalent to each other, and can't be discriminated based on the information in a single infinitely long two-state trajectory.

---

[1] The *on-off* correlation function as obtained from the trajectory is the same as the signal obtained from bulk experiments. This result is known as the Onsager principle [L. Onsager, Phys. Rev. **38**, 2265 (1931)].

In this paper, we present canonical forms of reduced dimensions (RD) [32, 33], e.g. Fig. 1C. RD forms have many advantageous over other canonical forms that were previously suggested in solving the problem of relating a mechanism to the time trajectory (see Refs. [31, 35]). RD forms can handle *any* KS, i.e. KSs with irreversible connections and/or symmetry also have RD form representation[2]. RD forms constitute a powerful tool in discriminating among KSs, because the mapping of a KS onto a RD form is based on the KS's *on-off* connectivity and therefore can be performed, to a large extent, without actual calculations. Using this property, we give an ensemble of relationships between properties of the data, the topology of the RD forms, and properties of KSs. These relationships are useful in discriminating KSs and in the analysis of the data.

Yet, finding the most reliable RD form from a finite trajectory is a real challenge. Here, we present the procedure for reliably constructing the RD form from finite data. RD forms can be constructed from data sets fairly accurately, and, importantly, much more accurately than other mechanisms (at least in the cases we studied, which represent commonly encountered mechanisms from such trajectories.)

The organization of the paper obeys: Part II presents RD forms and relates RD forms to the data and to KSs. In this part, the mathematical formulations of the system are presented. Part III builds a RD form from finite data. In this part, the statistical treatment of the analysis of finite length data is presented. Part IV concludes.

---

[2] Symmetry means that the spectrum of the waiting time probability density function (WT-PDF) for the single *x* (=*on, off*) periods is degenerate.

## II. RD forms and their relations to the data and to *on-off* KSs

In this part, we present the canonical forms of reduced dimensions (RD). We then relate the data and RD forms, and RD forms and KSs. Mathematical descriptions of RD forms, KSs and the data are presented as part of the discussion in this paragraph.

### A. RD forms

Here, we give all the information about RD forms. The mathematical proofs are given in **II.C** and in appendix **B**.

**Mathematical formulation of the data** For the following discussion, it is important to express the WT-PDFs for following events (i.e. *x* event followed by *y* event), $\phi_{x,y}(t_1,t_2)$s, as an expansion of exponentials. The WT-PDF $\phi_{x,y}(t_1,t_2)$ is built from the data by constructing the histogram of the intersection of successive *x* followed by *y* events. The most general mathematical description of the WT-PDF $\phi_{x,y}(t_1,t_2)$ that is constructed from a two-state trajectory generated by a KS follows the formula:

$$\phi_{x,y}(t_1,t_2) = \sum_{i=1}^{L_x}\sum_{j=1}^{L_y} \sigma_{x,y,ij} e^{-\lambda_{x,i}t_1 - \lambda_{y,j}t_2} . \tag{1}$$

Here, we use the matrix of amplitudes, $\sigma_{x,y}$, the sets of rates, $\{\lambda_x\}$ and $\{\lambda_y\}$, and the expansion lengths, $L_x$, and $L_y$. From Eq. (1), we can construct almost any quantity of interest. For example, integrating over $t_2$ leads to,

$$\phi_x(t) = \sum_{i=1}^{L_x} c_{x,i} e^{-\lambda_{x,i}t} , \tag{2}$$

with, $c_{x,i} = \sum_{j=1}^{L_y} \sigma_{x,y,ij}/\lambda_{y,j}$. Here, $\phi_x(t)$ is the WT-PDF of the *x* durations in the data. Note that the above amplitudes and rates can be expressed as a function of the kinetic rates of the underlying KS. (See **II.C**.) These expressions connect RD forms to KSs.

To understand RD forms and their relations to the data, the above equations are enough because RD forms are straightforwardly built from the matrices $\sigma_{x,y}$, and the rates, $\{\lambda_x\}$ and $\{\lambda_y\}$.

**Description of RD forms** The RD forms are networks made of substates. Each substate belongs to either the *on* state or the *off* state. Substates of the same states are never connected in RD forms. The connections are only between substates of different states. See Figs. 2D, 2G for examples.

A connection between substates in the RD form is always directional, and in RD forms it is possible to have substates that are connected only in one direction. A connection in a RD form represents a waiting time PDF that is almost always multi-exponential. We denote by $\varphi_{x,ij}(t)$ the WT-PDF in the RD form that connects substates $j_x \rightarrow i_y$. $\varphi_{x,ij}(t)$ is a sum of exponentials with as many components as in $\phi_x(t)$. In particular, in $\varphi_{x,ji}(t)$ the exponentials have rates $\{\lambda_x\}$ and amplitudes $\{\alpha_x\}$,

$$\varphi_{x,ji}(t) = \sum_{k=1}^{L_x} \alpha_{x,jki} e^{-\lambda_{x,k} t} . \qquad (3)$$

It is straightforward to get the amplitudes, $\alpha_{x,jki}$s, and the rates numerically when mapping a KS into a RD form (32-33). (The mapping of *on-off* KSs into RD forms is based on the path representation of the $\phi_{x,y}(t_1,t_2)$s as shown in part **II.C** and appendix **B**.) Estimating the amplitudes and rates from the data is a much harder task. Our toolbox presented in part **III** is designed to construct the RD form from finite binary data, and thus presents the way to construct the $\{\lambda_x\}$ and the $\{\alpha_x\}$ from the data.

**Examples of RD forms** The simplest topology for a RD form (2D) has one substate in each of the states. Therefore, $\varphi_{x,11}(t) = \phi_x(t)$. For a two by two RD form (2G), there are as many as 4 different $\varphi_{x,ji}(t)$s for each value of *x*. In general, for a RD form with $L_{RD,x}$ substates in state *x*, there are as many as $2L_{RD,on}L_{RD,off}$ different WT-PDFs for the connections in the RD form.

**Some basic properties of RD forms** A RD form has the minimal number of substates needed to reproduce the data. The simplicity in the topology of the RD form and the fact that only substates of different states are connected in the RD form cost a price: in RD forms each connection has a multi exponential WT-PDF. This means that RD forms are not Markovian mechanisms. (In contrast, KSs are Markovian mechanisms. Every KS has only single exponential WT-PDFs for the connections; however, the topology of a KS is always complex relative to the equivalent RD form.)

Microscopic reversibility is an important property of physical systems, and must hold when there is no external force acting on the system. For kinetic schemes, the microscopic reversibility condition is translated into: (*a*) the KS must have only reversible connections, and (*b*) no net flux is allowed at steady state within any closed loop in the KS. Now, it was mentioned before that a RD form can have irreversible connections. However, as RD forms are not Markovian mechanisms, a RD form can preserve microscopic reversibility on the *on-off* level even when it has irreversible connections. These can be 'balanced' by the existence of multi-component WT-PDFs for the connections. Microscopic reversibility in a RD form means that the $\phi_{x,y}(t_1,t_2)$ s obtained when reading the two-state trajectory in the forward direction are the same as the corresponding $\phi_{x,y}(t_1,t_2)$ s obtained when reading the trajectory backwards[3]. Using matrix notation, microscopic reversibility means, $\phi_{x,y}(t_1,t_2) = [\phi_{y,x}(t_1,t_2)]^T$, where *T* stands for the transpose of a matrix. When this condition holds the RD form built from the trajectory must fulfill microscopic reversibility.

---

[3] This is a general test that checks whether the data fulfils microscopic reversibility and was used in Ref. 39 for drawing conclusions about properties of aggregated Markov chains.

**RD form as canonical forms** RD forms are canonical forms in the sense that only one RD form can be constructed from an infinitely long two-state trajectory, and this RD form contains all the information in the two-state trajectory. RD forms are canonical forms of KSs because a given KS is mapped to a unique RD form.

**Simulating trajectories using RD forms** The fact that RD forms are not Markovian mechanisms is also seen in the way a trajectory is generated from a RD form. To produce an *on-off* trajectory from a given RD form, we simulate a random walk in the RD form using a modified Gillespie Monte Carlo algorithm[4]. Each transition in the simulation happens in two steps. Assume the process starts at substate $i_x$. The first step chooses the destination of the next location and is determined by the weights of making a transition $i_x \rightarrow j_y$: $w_{j_y i_x} = \overline{\varphi}_{x, j_y i_x}(0) / \sum_{j'_y} \overline{\varphi}_{x, j'_y i_x}(0)$. Here, the overbar symbol stands for a Laplace transform of a function, $\bar{f}(s) = \int_0^\infty f(t) e^{-st} dt$. The second step in the simulation of a jump in the RD form uses the chosen $j_y$ from the first step, and draws a random time out of a normalized density, $\varphi_{x, j_y i_x}(t) / \overline{\varphi}_{x, j_y i_x}(0)$. The procedure is then repeated at the new location. It is much faster to generate a two-state trajectory using the RD form than simulating a random walk in the underlying KS.

**Basic relations between RD forms and the data** The topology of a RD form is determined by the ranks, $R_{x,y}$s, of the corresponding $\phi_{x,y}(t_1, t_2)$s. For discrete time, $\phi_{x,y}(t_1, t_2)$ is a matrix with a rank $R_{x,y}$. $R_{x,y}$ is, in fact, the rank of $\sigma_{x,y}$. $R_{x,y}$ for $x \neq y$ is the number of substates in state $y$ in the RD form. (This rule holds in most cases for which $R_{x,y}$, for $x \neq y$, is larger than the same event ranks, $R_{x,x}$, $x = on, off$.) Because the $R_{x,y}$s are obtained from the $\phi_{x,y}(t_1, t_2)$s without the need of finding thier actual functional forms, we say that the topology of the RD form is found from the data without fitting.

---

[4] The original work by Gillespie (D. T. Gillespie, J. Comput. Phys. **22**, 403, 1976) dealt with simulating Markovian dynamics.

The fact that the rank $R_{x,y}$s determine the topology of the RD form also gives a set of important relations between the RD forms and *on-off* KSs. This is shown in **II.B-II.C**.

### B. KSs and RD forms, and utilities of RD forms

We postulate above that RD forms are canonical forms of two-state trajectories, and that RD forms are canonical forms of KSs, as a given KS is mapped to a unique RD form. The mathematical discussion is given in **II.C**. Now, we can take these postulations and present them in an illustrative way: we take the space of kinetic schemes, i.e. the space of all possible *on-off* Markovian mechanisms, and divide it into boxes. Each box can contain many KSs. However, each box has a unique RD form representation. Thus, each box is defined by the $R_{x,y}$s. A box however has also other properties: the number of exponentials in $\phi_x(t)$, $L_x$, and the complexity of the $\varphi_{x,ji}(t)$ are also important (mathematically, the $R_{x,y}$, $\{\alpha_x\}$, and $\{\lambda_x\}$ determine the properties of a box).

Note that RD forms can represent underlying KSs with symmetry and irreversible connections because they are built from all four $R_{x,y}$s, and this means that the space of KSs is not restricted.

The division of the space of KSs into RD forms relates the two sorts of mechanisms, and suggests that a mapping between KSs and RD forms exists. Indeed, this is shown in what follows.

**Basic relation between KSs and RD forms** The basic relation between RD forms and *on-off* KSs is based upon the identification of the ranks $R_{x,y}$ of KSs. For a given *on-off* KS, the rank $R_{x,y}$ for $x \neq y$ is deduced by analyzing the *on-off* connectivity in the KS. The basic rule is that the rank $R_{x,y}$ for $x \neq y$ obeys,

$$R_{x,y} = \min(M_x, N_y). \tag{4}$$

Here, $N_x$ and $M_x$ are the numbers of initial and final substates in state *x* in the *on-off* KS, respectively. Namely, each event in state *x* starts at one of the $N_x$ initial substates, and terminates through one of the $M_x$ final substates. (Important to note here is that there are cases in which the rank obeys a different formula than the one given above. For example, in symmetric KSs the above formula may fail. See part II.C for an elaborate discussion.)

Based on the general rule for determining the topology of RD forms from the data (recall that $R_{x,y}$ for $x \neq y$, as deduced from finding the rank of $\phi_{x,y}(t_1, t_2)$, is the number of substates in state *y* in the RD form), we obtain a relation between non-symmetric *on-off* KSs and the topology of RD forms. This relation is made possible through the ranks $R_{x,y}$ s. Further discussion is given in what follows and in part II.C.

The WT-PDFs for the connections in the RD form, the $\varphi_{x,ji}(t)$ s, are determined using the mapping procedure of a KS into a RD form. The mapping is presented in the next paragraph.

**Mapping a KS into a RD form** The basic rule for the mapping of a KS into a RD form is based on *clustering some of the initial substates in the KS into substates in the RD form*, where initial substates in the KS that are not clustered are mapped to themselves. The rule that determines if an initial substate in the KS is clustered or mapped to itself is as follows[5]: initial substates in state $y$ in the KS that contribute to $R_{x,y}$ ($x \neq y$) are mapped to themselves. The initial substates that do not contribute to $R_{x,y}$ are clustered, where initial-$y$-state substates in a cluster are all connected to the same final-$x$-state substate that contributes to $R_{x,y}$. (When substate $m_x$ has a single exit-connection to substate $n_y$, which is its only entering-connection, substate $n_y$ is defined as the one contributing to the rank). For example, the KS in Fig. 2E is mapped into a RD form (Fig. 2G) when clustering substates $1_{off}$-$2_{off}$ into the RD form's substates $1_{off}$ where substate $3_{off}$ is mapped to itself giving rise to substate $2_{off}$ in the RD form, because only substate $3_{off}$ contributes to the rank $R_{on,off}$ among the *off* initial substates in the KS. In the *on* state in this KS, substate $1_{on}$ is mapped to itself into substate $1_{on}$ in the RD form, and substates $2_{on}$-$3_{on}$ are clustered into RD forms substate $2_{on}$. Only substate $1_{on}$ contribute to $R_{off,on}$ among the *on* initial substates.

The clustering procedure determines the coefficients in the exponential expansion of the $\varphi_{x,ij}(t)s$. (Technical details for obtaining these WT-PDFs given a KS are discussed in Appendix B). Note that the clustering procedure, and the fact that substates in the KS that are not initial ones or final ones do not affect the RD form's topology, reduce the KS dimensionality to that of the RD form.

**Using RD forms to distinguish between KSs** We have already stated that the simplest topology for a RD form has one substate in each of the states, namely, $R_{x,y} = 1$ ($x, y = on, off$). This immediately means that when analyzing a trajectory from a mechanism of rank one, the only possible choice for $\varphi_{x,11}(t)$ is $\phi_x(t)$ (Fig. 2D). Namely, for this case all the information in the data is contained in $\phi_{on}(t)$ and $\phi_{off}(t)$. Consequently, KSs with $R_{x,y} = 1$ ($x, y = on, off$) and the same $\phi_{on}(t)$ and $\phi_{off}(t)$ are

---

[5] This rule applies for non symmetric KSs.

*indistinguishable* by the analysis of the data[6]. Examples of such KSs are shown in Fig. 2A-2C. The generalization of the equivalence of KSs for any case is straightforward using RD forms. KSs with the same $R_{x,y}$s and the same WT-PDFs for the connections in the RD form cannot be distinguished. Indistinguishable KSs with $R_{x,y}=2$ ($x,y=on,off$) and tri-exponential $\phi_{on}(t)$ and $\phi_{off}(t)$ are shown in Figs. 2E-2F, and their corresponding RD form is shown in Fig. 2G.

Clearly, two KSs with different $R_{x,y}$s can be resolved by the analysis of a two-state trajectory. Among the advantages of RD forms is to provide a powerful tool in resolving KSs with the same $R_{x,y}$s, and the same number of exponentials in $\phi_{on}(t)$ and $\phi_{off}(t)$, even without the need of performing actual calculations. This is done based only on *distinct complexity* of the WT-PDFs for the connections in the corresponding RD forms, e.g. compare the KSs in Fig. 3A and Fig. 3B, or, on different connectivity of RD forms, e.g. compare KSs in Figs. 3A-3B with the KS in Fig. 3C. The above means that it is impossible to find positive (> 0) kinetic rates for the KSs in Figs. 3A-3C that make the $\phi_{x,y}(t_1,t_2)$s from these KSs the same, so these KSs can be distinguished by analyzing a two-state trajectory (excluding symmetric cases for which the $\phi_{x,y}(t_1,t_2)$s factorizes to the product of $\phi_x(t_1)\phi_y(t_2)$s).

**Trajectories as a function of an external parameter** The division of KSs into RD forms is useful also when on top of the information extracted from the 'original' trajectory, complementary details about the observed process are available.

Complementary details are obtained from analyzing different types of measurements of the system, e.g. the crystal structure of the biopolymer, or from analyzing two-state trajectories while varying some parameters, e.g. the substrate concentration [13-15].

---

[6] This statement refers to a situation when additional information on the underlying mechanism is known.

Suppose that the connectivity of the underlying KS is unchanged by the manipulation. Then, the additional information can be used to resolve KSs that correspond to the RD form found from the statistical analysis of the 'original' two-state trajectory, whereas any KS with a different RD form is irrelevant. Alternatively, when manipulating the system leads to a change in the connectivity of the underlying KS, or even to the addition or removal of substates, the RD forms obtained from the different data sets are distinct. RD forms can handle both scenarios; in the first case an adequate parameter tuning relates the RD forms obtained from the various sources, whereas in the second case the RD forms cannot be related by a parameter tuning.

**Summary of the utilities of RD forms** Let us summarize the basic utilities of RD forms: (*a*) A RD form has the simplest topology that can reproduce the data. (*b*) The topology of the RD form is obtained from the data without fitting. (*c*) RD forms can represent KSs with symmetry and irreversible transitions because these canonical forms are built from all four $R_{x,y}$s. (*d*) RD forms constitute a convenient and powerful tool, a much more powerful tool than other methods, for discriminating among *on-off* KSs, which means determining whether two different *on-off* KSs lead to the same data in a statistical sense (namely, to the same RD form). Important to note here is that when collecting data while changing an external parameter equivalent KSs may be distinguished. This is possible when assuming particular dependencies of the kinetic rates on the external parameter. Conceptually, the data in such a case is no longer a two-state data, but made of additional dimensions.

The above points show the strength of RD forms in dealing with the problem theoretically. Although, these points can't be used directly to construct the RD form from realistic data, these points are also convenient features to rely on when constructing the RD form from the data. For example, property (*a*) above guarantees that RD form has the simplest topology that can generate the data. Property (*b*) guarantees that the topology is found from the data without fitting. Both features facilitate the construction of the RD form from the data. Part **III** presents the way to construct RD forms from realistic data, and shows in a more precise way how these properties are indeed useful in such a construction[7].

## C. Mathematical formulations and proofs for the RD forms' connections to the data and to KSs

In this part, we express the WT-PDFs for single periods, $\phi_x(t)$, $x = on, off$, and for joint successive periods, $\phi_{x,y}(t_1,t_2)$, $x, y = on, off$, in terms of both the master equation and the path representation. *On-off* KSs are commonly described in terms of the master equation, whereas our canonical forms are naturally related to the path representation. The relationship between the two representations is made, and this establishes the mathematical relation between *on-off* KSs and RD forms.

**Matrix formulation of the system** The random walk in an *on-off* KS is formulated using a master equation. The time-dependent occupancy probabilities of state *x* for the coupled *on-off* process, $\vec{P}_x(t)$, with the elements $\left(\vec{P}_x(t)\right)_i = P_{x,i}(t)$ for $i = 1,...,L_x$ ($L_x$ is the number of substates in state *x*), obey the reversible master equation,

$$\frac{\partial}{\partial t}\begin{pmatrix}\vec{P}_{on}(t)\\ \vec{P}_{off}(t)\end{pmatrix} = \begin{pmatrix}\mathbf{K}_{on} & \mathbf{V}_{off}\\ \mathbf{V}_{on} & \mathbf{K}_{off}\end{pmatrix}\begin{pmatrix}\vec{P}_{on}(t)\\ \vec{P}_{off}(t)\end{pmatrix}. \quad (5)$$

---

[7] The utilities of RD forms makes RD forms the most reliable mechanism to construct from the data relative to other mechanisms, but also, makes it strong in an absolute way. (This is to say that the obtained RD form is very close to the one that generated the data. See part **III** for information.)

In Eq. (5), the matrix $\mathbf{K}_x$, with dimensions $[\mathbf{K}_x] = L_x, L_x$, contains kinetic rates among substates in state $x$, and 'irreversible' kinetic rates from substates in state $x$ to substates in state $y$. (The 'irreversible' kinetic rates appear on the diagonal, and are called irreversible because matrix $\mathbf{K}_x$ does not contain the back kinetic rates from state $y$ to state $x$). Matrix $\mathbf{V}_x$, with dimensions $[\mathbf{V}_x] = L_y, L_x$, contains kinetic rates between states $x \to y$, where $(\mathbf{V}_x)_{ji}$ is the kinetic rate between substates, $i_x \to j_y$. $\vec{\mathbf{P}}_x(ss)$ is the vector of occupancy probabilities in state $x$ at steady state ($t \to \infty$). $\vec{\mathbf{P}}_x(ss)$ is found from Eq. (5) for vanishing time derivative. $\phi_x(t)$ and $\phi_{x,y}(t_1, t_2)$ are given in terms of the matrices in Eq. (5),

$$\phi_x(t) = \vec{\mathbf{1}}_y^T \mathbf{V}_x \mathbf{G}_x(t) \mathbf{V}_y \vec{\mathbf{P}}_y(ss) / N_x, \tag{6}$$

and,

$$\phi_{x,y}(t_1, t_2) = \vec{\mathbf{1}}_x^T \mathbf{V}_y \mathbf{G}_y(t_2) \mathbf{V}_x \mathbf{G}_x(t_1) \mathbf{V}_y \vec{\mathbf{P}}_y(ss) / N_x, \tag{7}$$

where $N_x = \vec{\mathbf{1}}_x^T \mathbf{V}_y \vec{\mathbf{P}}_y(ss)$ and $\vec{\mathbf{1}}_x^T$ is the summation row vector of $1, L_x$ dimensions, $[\vec{\mathbf{1}}_x^T] = 1, L_x$. The expression for $\phi_{x,x}(t_1, t_2)$ is obtained from Eq. (7) when plugging in the factor $\mathbf{V}_y \overline{\mathbf{G}}_y(0)$,

$$\phi_{x,x}(t_1, t_2) = \vec{\mathbf{1}}_y^T \mathbf{V}_x \mathbf{G}_x(t_2) \mathbf{V}_y \overline{\mathbf{G}}_y(0) \mathbf{V}_x \mathbf{G}_x(t_1) \mathbf{V}_y \vec{\mathbf{P}}_y(ss) / N_x.$$

Here, $\overline{\mathbf{G}}_y(0)$ is the Laplace transform of $\mathbf{G}_y(t)$, $\overline{\mathbf{G}}_y(s) = \int_0^\infty \mathbf{G}_y(t) e^{-st} dt$, at $s=0$. $\mathbf{G}_x(t)$ in the above equations is the Green's function of state $x$ for the irreversible process, $\partial \mathbf{G}_x(t) / \partial t = \mathbf{K}_x \mathbf{G}_x(t)$, with the solution,

$$\mathbf{G}_x(t) = \exp(\mathbf{K}_x t) = \mathbf{X} \exp(\lambda_x t) \mathbf{X}^{-1}. \tag{8}$$

The second equality in Eq. (8) follows from a similarity transformation $\boldsymbol{\lambda}_x = \mathbf{X}^{-1}\mathbf{K}_x\mathbf{X}$, and all the matrices in Eq. (8) have dimensions $L_x, L_x$. Non symmetric KSs must have non-degenerate eigenvalue matrices $\boldsymbol{\lambda}_x$s.

This paragraph showed that given a KS, it is very simple to build any observable that can be constructed from the measurement; in particular, the WT-PDFs for a single event or two subsequent events, Eqs. (6-8).

**Path representation of the WT-PDFs** Our approach, from which RD forms naturally emerge, is based on expressing the WT-PDFs in Eqs. (6)-(7) in path representation that utilizes the *on-off* connectivity of the KS. The *on-off* process is separated into two irreversible processes that occur one after the other, and we have for $\phi_{x,y}(t_1,t_2)$ ($x \neq y$),

$$\phi_{x,y}(t_1,t_2) = \sum_{n_y=1}^{N_y}\left(\sum_{n_x=1}^{N_x}W_{n_x}f_{n_yn_x}(t_1)\right)F_{n_y}(t_2) = \sum_{m_x \in \{M_x\}}\left(\sum_{n_x=1}^{N_x}\sum_{n_y=1}^{N_y}W_{n_x}\tilde{f}_{m_xn_x}(t_1)\omega_{n_ym_x}F_{n_y}(t_2)\right). \quad (9)$$

(In Appendix A expressions for $\phi_x(t)$ and $\phi_{x,x}(t_1,t_2)$ in path representation are given). Here, a sum $z_x \in \{Z_x\}$ is a sum over a particular group of $Z_x$ substates. Equation (9) emphasizes the role of the KS's topology in expressing the $\phi_{x,y}(t_1,t_2)$ s. $N_x$ and $M_x$ are the numbers of initial and final substates in state $x$ in the KS, respectively. Namely, each event in state $x$ starts at one of the $N_x$ initial substates, labeled, $n_x = 1,..., N_x$, and terminates through one of the $M_x$ final substates, labeled $m_x = 1,…, M_x$, for a reversible *on-off* connection KS (Fig. 2E), or $m_x = N_x +1-H_x,…, N_x +M_x-H_x$, for an irreversible *on-off* connection KS (Fig. 1B), where $H_x$ (= 0,1,…, $N_x$) is the number of substates in state $x$ that are both initial and final ones. (In each of the states the labeling of the substates starts from 1). An event in state $x$ starts in substate $n_x$ with probability $W_{n_x}$. The first passage time PDF for exiting to substate $n_y$, conditional on starting in substate $n_x$ ($x \neq y$), is $f_{n_y n_x}(t)$, and $F_{n_x}(t) = \sum_{n_y} f_{n_y n_x}(t)$. Writing $f_{n_y n_x}(t)$ as, $f_{n_y n_x}(t) = \sum_{m_x} \omega_{n_y m_x} \tilde{f}_{m_x n_x}(t)$, emphasizes the role of the *on-off* connectivity in the KS for expressing $f_{n_y n_x}(t)$, where $\omega_{n_y m_x}$ is the transition probability from substate $m_x$ to substate $n_y$, and $\tilde{f}_{m_x n_x}(t) \omega_{n_y m_x}$ is the first passage time PDF, conditional on starting in substate $n_x$, for exiting to substate $n_y$ through substate $m_x$.

When analyzing Eq. (9), we see that it can be expressed as a sum of rank-one 2D WT-PDFs. The exact number of such rank-one 2D WT-PDFs in the sum obeys Eq. (4). This is the first time that we can see a mathematical connection between KSs and RD forms. Indeed RD forms are based (mathematically) on Eq. (9) and its generalization [see Eq. (10)].

**Relationships between the master equation and the path representation** All the factors in Eq. (9) can be expressed in terms of the matrices of Eq. (5). $W_{n_x}$ and $f_{n_y n_x}(t)$ are related to the master equation through the following relation:

$$W_{n_x} = \left(\mathbf{V}_y \vec{\mathbf{P}}_y(ss)\right)_{n_x} / N_x,$$

and,

$$f_{n_y n_x}(t) = \left(\mathbf{V}_x \mathbf{G}_x(t)\right)_{n_y n_x}.$$

$f_{n_y n_x}(t)$ can be rewritten as,

$$f_{n_y n_x}(t) = \sum_{m_x} \omega_{n_y m_x} \tilde{f}_{m_x n_x}(t),$$

and similarly for $\left(\mathbf{V}_x \mathbf{G}_x(t)\right)_{n_y n_x}$ we have,

$$\left(\mathbf{V}_x \mathbf{G}_x(t)\right)_{n_y n_x} = \sum_k \left(\mathbf{V}_x\right)_{n_y k} \left(\mathbf{G}_x(t)\right)_{k n_x}.$$

Note that the factors in the right hand side in the above two sums are not equal but proportional,

$$\tilde{f}_{k n_x}(t) = \tilde{\alpha}_{x,k} \left(\mathbf{G}_x(t)\right)_{k n_x} \quad ; \quad \tilde{\alpha}_{x,k} = -\left(\mathbf{K}_x\right)_{kk},$$

and

$$\omega_{n_y k} = \left(\mathbf{V}_x\right)_{n_y k} / \tilde{\alpha}_{x,k}.$$

The fact that there is a simple connection between the master equation and the path representation means that there is a connection between RD forms and KSs. Recall that it is the path representation which sets the basis for the mapping of a KS into a RD form (the mapping that was spelled out in the previous part, **II.B**).

**The rank of $\phi_{x,y}(t_1, t_2)$ and its topological interpretation** Some of the discussion in this paragraph was presented before, but here, we include mathematical support. Recall that as time is always discrete (in applications), $\phi_{x,y}(t_1, t_2)$ is a matrix. Therefore, it has a rank, $R_{x,y}$ (= 1, 2, …). $R_{x,y}$ is the number of non-zero eigenvalues (or singular values for a non square matrix) in the decomposition of $\phi_{x,y}(t_1, t_2)$, and so can be obtained without the need of finding the actual functional form of $\phi_{x,y}(t_1, t_2)$. Using Eq. (9), which gives $\phi_{x,y}(t_1, t_2)$ as sums of terms each of which is a product of a function of $t_1$ and a function of $t_2$, we can relate $R_{x,y}$ ($x \neq y$) to the topology of the underlying KS. When none of the terms in an external sum on Eq. (9), after the first or the second equality, are proportional,

$$R_{x,y} = \min(M_x, N_y) \tag{4.1}$$

(see Figs. 2A-2C, and Fig. 2F). Otherwise,

$$R_{x,y} < \min(M_x, N_y) \tag{4.2}$$

(see Fig. 2E, and Appendix B), and Eq. (9) is rewritten such that it has the *minimal* number of additives in the external summations,

$$\phi_{x,y}(t_1,t_2) = \sum_{n_y \in \{\tilde{N}_y\}} \left( \sum_{n_x=1}^{N_x} W_{n_x} f_{n_y n_x}(t_1) \right) F_{n_y}(t_2)$$

$$+ \sum_{m_x \in \{\tilde{M}_x\}} \left( \sum_{n_x=1}^{N_x} W_{n_x} \tilde{f}_{m_x n_x}(t_1) \right) \left( \sum_{n_y \notin \{\tilde{N}_y\}} \omega_{n_y m_x} F_{n_y}(t_2) \right). \quad (10)$$

This leads to the equality,

$$R_{x,y} = \tilde{N}_y + \tilde{M}_x. \quad (11)$$

$\tilde{N}_y$ and $\tilde{M}_x$ can be related to the KS's *on-off* connectivity. Consider a case where $M_x < N_y$, and there is a group of final substates in state $x$, $\{O_x\}$, with connections *only* to a group of initial substates in state $y$, $\{O_y\}$, and $O_x > O_y$ (see Fig. B4 in Appendix B). Then $\tilde{M}_x = M_x - O_x$ and $\tilde{N}_y = O_y$. (Further discussion and a generalization of this relationship are given in Appendix B).

This paragraph and the discussion in appendix B give the mathematical basis for the mapping of a KS into RD form. The fact that there is a single mapping between KSs and RD forms means that RD forms are canonical forms of *on-off* KSs. RD forms are canonical forms of two-state trajectories as they are built from the data uniquely (only one RD form can be built from an ideal data and this RD form contains all the information in the data.)

### III. Constructing the RD form from the data

So far, we have shown that RD forms are reliable mechanisms to work with in analyzing and dealing with time trajectories with discrete states. The question now is: can't a RD form be constructed reliably from the realistic data? In this part of this paper, we show that this is possible; namely, we present a way to construct the RD form from a finite noiseless trajectory. This makes RD forms a strong tool in analyzing the data also in practice.

Let us start and present an algorithm that builds a RD form from a finite data. This algorithm is based on the following three steps:

(*1*) Obtain the spectrum of the $\phi_x(t)$s by the Padé approximation [57]. The spectrum of the WT-PDFs for the *x* to *y* connections in the RD form is the same spectrum as that of $\phi_x(t)$, because substates of the same state in the RD form are not connected. Differences between $\phi_x(t)$ and the WT-PDFs in the set $\{\varphi_x(t)\}$ lay in the pre-exponential coefficients.

(*2*) Find the ranks of the $\phi_{x,y}(t_1,t_2)$ s, and use it to build the RD form topology.

(*3*) Apply a maximum likelihood procedure for finding the pre-exponential coefficients of the $\{\varphi_x(t)\}$. Our routine for maximizing the likelihood function uses its analytical derivatives.

We use the above three-stage algorithm in the construction of the RD form in Fig. 1C from finite data. This is a very simple test just to show that it is indeed possible to construct the RD form from the data. More complicated tests are presented in [33], and in work we are performing these very recent days to be published elsewhere.

Now, the RD form in 1C is a mapping of the KS in Fig. 1B. A two-state trajectory with one million cycles is produced (part of the trajectory is shown in Fig. 1A) when simulating a random walk in the RD form. The simulations are based upon a modified Gillespie Monte-Carlo method as explained in part II.A.

Figure 4 displays the analytical and the experimental $\phi_{on}(t)$ and $\phi_{off}(t)$. The $\phi_x(t)$ s are accurately obtained from the data for times such that their amplitudes are 2 orders of magnitude smaller than their maximal values. The Padé approximation method gives the correct amplitudes, rates, and number of components for these WT-PDFs.

The next stage in the construction of RD form estimates the ranks of the $\phi_{x,y}(t_1,t_2)$ s. Clearly, when working with the analytical $\phi_{x,y}(t_1,t_2)$ s, the singular value decomposition (SVD) method gives the correct values for the $R_{x,y}$ s, $R_{x,y}=2$ $\forall x,y=on,off$, but the ratio of the larger to smaller singular value is large (~$10^3$). This result means that the contribution from the large singular value contains most of the signal, and corresponds to the limit of an infinitely long trajectory. Thus, one may expect technical difficulties in detecting the exact number of nonzero singular values from an experimental matrix, due to the limited number of events in the trajectory. To deal with this issue, we build a series of cumulative 2D WT-PDFs. The first order cumulative of $\phi_{x,y}(t_1,t_2)$ is defined by,

$$c_1\phi_{x,y}(T_1,T_2) = \int_0^{T_1}\int_0^{T_2} dt_1 dt_2 \phi_{x,y}(t_1,t_2),$$

and the generalization to higher order cumulative PDFs naturally follows,

$$c_n\phi_{x,y}(T_1,T_2) = \int_0^{T_1}\int_0^{T_2} dt_1 dt_2 c_{n-1}\phi_{x,y}(t_1,t_2).$$

A cumulative two dimensional PDF reduces the noise in the *original* PDF, but also preserves the rank of the original PDF. For each two dimensional PDF we obtain its spectrum of singular values and plot the ratio of successive singular values as a function of the order of the large singular value in the ratio. This plot should show large values for signal ratios, and a constant behavior with a value of about a unity for noise ratios. Figures 5A-5B shows the singular value ratio method applied on $\phi_{on,off}(t_1,t_2)$ and its first three cumulative PDFs. Both the second and the third cumulative PDFs show large values for the first two ratios and a constant behavior for larger ratios. This is a signature for a rank two histogram. (Note that this behavior is not seen in the original PDF and it's first cumulative.) For comparison, Fig. 5C plots the same quantities calculated for $\phi_{off,on}(t_1,t_2)$, all of which show the same constant behavior for ratios greater than one, indicating a rank one histogram. A rank one behavior is observed also for $\phi_{on,on}(t_1,t_2)$ and $\phi_{off,off}(t_1,t_2)$ (data not shown).

At this stage, we can build a low resolution RD form with one *on* substate, because $R_{off,on}=1$, and two *off* substates, because $R_{on,off}=2$. Then, a search for 8 parameters, denoted by $\Theta = \{\alpha_{x,jHi}\}$, is performed. These are the parameters that appear in the exponential expansions of the $\varphi_{x,ij}(t)$ s,

$$\varphi_{x,ji}(t) = \sum_{H=1}^{2} \alpha_{x,jHi} e^{-\lambda_{x,H} t} . \tag{12}$$

The search for the best set $\Theta$ is performed using maximum likelihood calculations. (Note that some of the $\alpha_{x,jHi}$ s are zeros, but this information is not known a priori.) The likelihood function is given by,

$$L(\Theta | data) = \sum_{x,y} \sum_{i} \log(\phi_{x,y}(t_i, t_{1+i})).$$

The constraints in the algorithm demand that Eqs. (12) are positive for every value of *t*, and also that the normalization of the WT-PDFs for the connection,

$$\sum_{j} \int_{0}^{\infty} \varphi_{x,ji}(t)dt = 1,$$

holds for every *x* and *i*. The analytical derivatives of the likelihood function are used in the maximization procedure. The optimization procedure is performed in less than a minute for data made of $10^3$ events. Here, a straightforward optimization gave the correct answer[8]. For the studied case, *the error bars in the elements in {$\Theta$} are*

---

[8] Note that the optimization can also find local maximum, so, for a general case, it is required to choose various sets of initial conditions. In fact the problem of finding the best initial conditions for the maximization procedures is serious, as the maximization always find local minimum when the parameter space contains more than several variables. The complexity of the problem increases with increasing of the parameter space. In [33], we have developed a way to find the best set of initial

*always less than a percent of the actual values*. (The error bars are found by inverting the Hessian matrix of second derivatives of the likelihood function with respect to the unknowns and substituting the solution for the unknowns. The diagonal elements of the obtained matrix give the variance of the fit [58].)

This simple example illustrated the way a RD form can be constructed in a reliable way from the data.

### IV. Concluding remarks

This paper deals with the problem of deducing a reliable and unique mechanism from a time trajectory with two states that originated from a very complicated mechanism, say, a multi-substate *on-off* KS. For this, we construct a unique mechanism of reduced dimensions from the data. Our reduced dimensions mechanisms are canonical forms. The way to think about the canonical forms is as follows: the *on-off* KS space is partitioned into RD forms. Many KSs can have the same RD form, however, a given trajectory, and a given KS, is equivalent to only a unique RD form.

RD forms are *on-off* networks. A RD form has connections only between substates of different states. The connections in the RD form are usually non-exponential WT-PDFs. This means that RD forms (in these cases, and thus, in principle) are not Markovian mechanisms. The topology of the RD form is determined by the ranks $R_{x,y}$s of the $\phi_{x,y}(t_1,t_2)$s: $R_{x,y}$ for $x$ different than $y$ is the number of substates in state $y$ in the RD form. (This is the basic rule. For special cases the rules that relate the ranks to the RD form topology are more involved.)

A set of relations between the ranks $R_{x,y}$s and the KS's *on-off* connectivity was given. These relations enable mapping a KS into a RD form based only on the *on-off*

---

conditions such that the search time is minimized, and the final results are good enough. Further developments are performed these days, and will be published elsewhere.

connectivity of the KS. The relations are based on the path representation of the $\phi_{x,y}(t_1,t_2)$ s presented in Eqs. (9)-(10).

Among the advantages of RD forms is that they constitute a powerful tool in discriminating among *on-off* KSs.

An example that builds a RD form from a data of $10^6$ *on-off* cycles was presented, showing the applicability of RD forms in analyzing realistic data.

Concluding: based on the above strong properties of RD forms in the context of analyzing discrete-state trajectories and the applicability of RD forms in analyzing realistic (finite) discrete-state trajectories, we envision RD forms as a very valuable and abundantly-used tool in analyzing measurements from individual complexes and macromolecules, and other related dynamical signals. Based on this vision, we have started recently to develop software that will enable one to find a mechanism (a RD form with several possible equivalent KSs) from a given trajectory, and also gives other properties of the dynamical signal independent of a mechanism.

We intend to start with the distribution of this software during early 2010.

## Appendix A

In this Appendix, we give expressions for $\phi_x(t)$ and $\phi_{x,x}(t_1,t_2)$ using the path representation. Here, and in Appendix B, $x \neq y$, unless otherwise is explicitly indicated. The expression for $\phi_x(t)$ is obtained from Eq. (9) by integrating over one time argument, $\phi_x(t) = \int_0^\infty \phi_{x,y}(t,\tau)d\tau = \int_0^\infty \phi_{y,x}(\tau,t)d\tau$, which leads to,

$$\phi_x(t) = \sum_{n_y=1}^{N_y}\sum_{n_x=1}^{N_x} W_{n_x} f_{n_y n_x}(t) = \sum_{m_x \in \{M_x\}} \sum_{n_x=1}^{N_x}\sum_{n_y=1}^{N_y} W_{n_x} \tilde{f}_{m_x n_x}(t) \omega_{n_y m_x} . \tag{A1}$$

The expression for $\phi_{x,x}(t_1,t_2)$ is obtained when introducing an additional summation that represents the random walk in state $y$ that takes place in between the two measured events in state $x$,

$$\phi_{x,x}(t_1,t_2) = \sum_{n'_x=1}^{N_x} \sum_{n_y=1}^{N_y} \sum_{n_x=1}^{N_x} W_{n_x} f_{n_y n_x}(t_1) p_{n'_x n_y} F_{n'_x}(t_2)$$

$$= \sum_{m_x \in \{M_x\}} \sum_{n'_x=1}^{N_x} \sum_{n_y=1}^{N_y} \sum_{n_x=1}^{N_x} W_{n_x} \tilde{f}_{m_x n_x}(t_1) \omega_{n_y m_x} p_{n'_x n_y} F_{n'_x}(t_2). \quad (A2)$$

Here, $p_{n'_x n_y}$ is the probability that an event that starts at substate $n_y$ exits to substate $n'_x$, and is given by $p_{n'_x n_y} = \bar{f}_{n'_x n_y}(0)$, where $\bar{g}(s) = \int_0^\infty g(t) e^{-st} dt$ is the Laplace transform of $g(t)$.

Note that higher order successive WT-PDFs e.g. $\phi_{x,y,z}(t_1,t_2,t_3)$, do not contain additional information on top of the $\phi_{x,y}(t_1,t_2)$ s. When the underlying KS has no symmetry (i.e. the spectrum of $\phi_x(t)$, $x = on, off$, is non-degenerate) and/or irreversible connections, it is sufficient to use $\phi_{x,y}(t_1,t_2)$ for $x \neq y$, where for other cases, $\phi_{x,x}(t_1,t_2)$ s, $x = on, off$, contain complementary information.

## Appendix B

In this Appendix, we give expressions for the WT-PDFs for the connections in the RD form, denoted by $\varphi_{x,ij}(t)$ s, for any KS. We do not consider symmetric KSs separately, because symmetry is apparent in the functional form of the $\varphi_{x,ij}(t)$ s. Further discussion regarding the topological interpretation of $\tilde{M}_x$ and $\tilde{N}_y$ in Eq. (11) is also given.

The waiting time PDFs for the connections in the RD form are uniquely determined by the clustering procedure in the mapping of a KS into a RD form. The clustering procedure is based upon the identification of substates in the *on-off* connectivity that contribute to the ranks $R_{x,y}$. The four ranks $R_{x,y}$s determine the RD form topology, and the mapping determine the incoming flux and outgoing flux for each substate in the RD form. This makes the RD forms legitimate canonical forms that preserve all the information contained in the two-state trajectory.

The technical details to obtain the $\varphi_{x,ij}(t)$ s, given a KS, are spelled out below when considering two cases: (*1*) None of the terms in an external sum in Eq. (9), after the first or second equality, are proportional to each other, and (*2*) Some of the terms in an external sum in Eq. (9), after the first or second equality, are proportional to each other.

(*1.1*) *Reversible on-off connection KSs* Say, $M_x \geq N_y$, or equivalently $N_x \geq M_y$. (Fig. B1A with *x=off*). Based on the clustering procedure, there are $N_y$ substates in each of the states in the RD form, and as many as $2N_y^2$ WT-PDFs for the connections in the RD form. Initial substates in state *x* are clustered, and the expression for $\varphi_{x,n_y i_x}(t)$ reads,

$$\varphi_{x,n_y i_x}(t) = \frac{1}{N_{x,m_y}} \sum_{n_x} P_{y,m_y}(ss)(\mathbf{V_y})_{n_x m_y} f_{n_y n_x}(t). \tag{B1}$$

In Eq. (B1), we use the normalization $N_{x,m_y}$, defined through the equations,

$$N_x = \vec{\mathbf{1}}_x^T \mathbf{V}_y \vec{\mathbf{P}}_y(ss) = \sum_{m_y, n_x} P_{y,m_y}(ss)(\mathbf{V_y})_{n_x m_y} = \sum_{m_y} N_{x,m_y} = \sum_{n_x} N_{x,n_x}.$$

In Eq. (B1), we set $j_y \to n_y$ because there are $n_y = 1,...,N_y$ substates in state *y* in the RD form, and we can also employ the meaning of $n_y$ as the initial substates in state *y* in the underlying KS. Additionally, we associate $m_y$ on the right hand side (RHS), which has the meaning of final substates in the underlying KS, with $i_x$ on the left hand side (LHS), i.e. $m_y \to i_x$. Note that for a KS with only reversible *on-off* connections, $m_y = 1,...,M_y$, so the values of $m_y$ and $i_x$ can be the same.

The expression for $\varphi_{y,i_x n_y}(t)$ is different than that for $\varphi_{x,n_y i_x}(t)$ in both the normalization used and the factors that are summed, which is a result of the mapping of the initial substates in state *y* to themselves. $\varphi_{y,i_x n_y}(t)$ is given by,

$$\varphi_{y,i_x n_y}(t) = \frac{1}{N_{y,n_y}} \sum_{m_x} P_{x,m_x}(ss)(\mathbf{V_x})_{n_y m_x} \tilde{f}_{m_y n_y}(t) \tilde{\omega}_{m_y} \quad ; \quad \tilde{\omega}_{m_y} = \sum_{n_x} \omega_{n_x m_y} . \quad (B2)$$

Note that here, $\varphi_{y,i_x n_y}(t) = \tilde{f}_{m_y n_y}(t)\tilde{\omega}_{m_y} = (\mathbf{G_y}(t))_{m_y n_y} \sum_{n_x} (\mathbf{V_y})_{n_x m_y}$. In Eq. (B2), we associate $m_y$ on the RHS with $i_x$ on the LHS, i.e. $m_y \to i_x$. Again, for a KS with only reversible transitions, the $i_x$s can have the same values as of the $m_y$s.

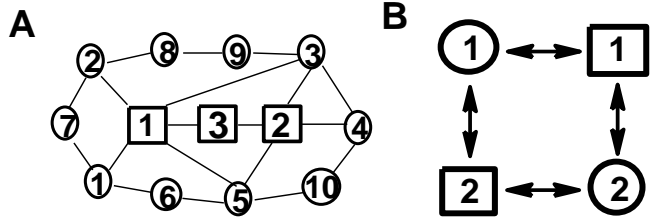

**FIG B1 A** – A reversible connection KS, with $N_{on}=M_{on}=2$ and $N_{off}=M_{off}=5$. **B** - The RD form of KS **A**. The RD form's substate $1_{off}$ corresponds to the cluster of the KS's *off* substates $1_{off}$-$3_{off}$ and $5_{off}$, because these are connected to substate $1_{on}$ in the KS, which contributes to the rank $R_{on,off}$. The RD form's substate $2_{off}$ corresponds to the cluster of the KS's *off* substates $3_{off}$-$5_{off}$, because these are connected to substate $2_{on}$ in the KS, which contributes to the rank $R_{on,off}$. Note that a particular initial substate can appear in more than a single cluster, which simply means that the overall steady-state flux into the substate in divided into several contributions. The initial *on* substates in the KS both contribute to $R_{off,on}$ so they are mapped to themselves in the RD form. The WT-PDFs for the connections can be obtained from Eqs. (B1)-(B2).

(*1.2*) *Irreversible on-off connection KSs* Obtaining the $\varphi_{x,ij}(t)$s for irreversible *on-off* connection KSs is similar to getting these WT-PDFs for reversible *on-off* connection KSs. The reason is that the clustering procedure is based on the *directional* connections between final substates in state *x* and initial substates in state *y*. However, some technical details may differ. We consider two cases.

(*1.2.1*) Let $M_x \geq N_y$ and $M_y \geq N_x$. (Fig. B2A-B2B). Then, the WT-PDFs for the connections are given by,

$$\varphi_{x,n_y n_x}(t) = \frac{1}{N_{x,n_x}} \sum_{m_y} P_{y,m_y}(ss)(\mathbf{V_y})_{n_x m_y} f_{n_y n_x}(t) = f_{n_y n_x}(t), \qquad (B3)$$

and,

$$\varphi_{x,n_x n_y}(t) = \frac{1}{N_{y,n_y}} \sum_{m_x} P_{x,m_x}(ss)(\mathbf{V_x})_{n_y m_x} f_{n_x n_y}(t) = f_{n_x n_y}(t). \qquad (B4)$$

Note that for this case any $\varphi_{z,ij}(t)$ equal to the corresponding $f_{ij}(t)$. This is an outcome of the KS's topology for which in both the *on* to *off* and the *off* to *on* connections, the number of initial substates in a given state is lower than the number of final substates in the other state.

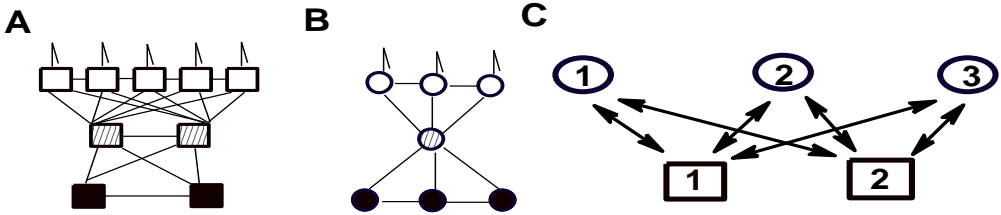

**FIG B2** An example for a KS with irreversible *on-off* connections, and $N_{on}=2$, $M_{on}=5$, $N_{off}=3$, and $M_{off}=3$. The KS is divided into two parts shown on **A** (*on* state) and **B** (*off* state) for a convenient illustration. The RD form is shown on **C**. The WT-PDFs for the connections can be obtained from Eqs. (B3) - (B4).

(*1.2.2*) Let $N_x > M_y$ and $N_y > M_x$. (Fig. B3A-B3B). Then, the WT-PDFs for the connections are given by,

$$\varphi_{x,j_y i_x}(t) = \frac{1}{N_{x,m_y}} \sum_{n_x} P_{y,m_y}(ss)(\mathbf{V_y})_{n_x m_y} \tilde{f}_{m_x n_x}(t) \tilde{\omega}_{m_x}, \qquad (B5)$$

and,

$$\varphi_{y,i_x j_y}(t) = \frac{1}{N_{y,m_x}} \sum_{n_y} P_{x,m_x}(ss)(\mathbf{V_x})_{n_y m_x} \tilde{f}_{m_y n_y}(t) \tilde{\omega}_{m_y}. \qquad (B6)$$

In Eqs. (B5)-(B6), we use the mapping $m_y \to i_x$ and $m_x \to j_y$ between the RHS and the LHS indexes. (In particular, $m_y - (N_y - H_y) = i_x$, and $m_x - (N_x - H_x) = j_y$).

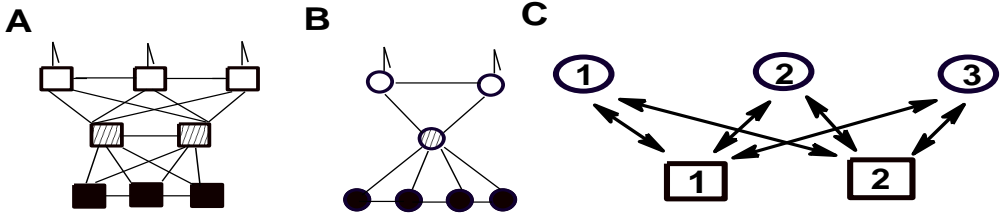

**FIG B3** An irreversible *on-off* connection KS with $N_{on}=3$, $M_{on}=3$, $N_{off}=4$, and $M_{off}=2$. The panels are divided as in Fig. B2. The WT-PDFs for the connections can be obtained from Eqs. (B5) - (B6).

(*2*) We turn now to deal with cases in which some of the terms in Eq. (9) are proportional, and therefore Eq.(10) is used for expressing $\phi_{x,y}(t_1,t_2)$. We consider only KSs with reversible *on-off* connections, but the same analysis is releavnt for KSs with irreversible *on-off* connections.

Let $M_x \leq N_y$, or equivalently $N_x \leq M_y$. (See Fig. B4A with $x=off$). So it follows that, $R_{x,y} < M_x$, which is a result of a special *on-off* connectivity. In particular, let $\{O_y\}$ and $\{O_x\}$ be the groups of substates in states $y$ and $x$ respectively, such that the substates in $\{O_x\}$ are connected only to the substates in $\{O_y\}$, and $O_y < O_x$. (In Fig. B4A, the group $\{O_{off}\}$ contains the substates $1_{off}$, $2_{off}$ and $3_{off}$, and the group $\{O_{on}\}$ contains the substates $1_{on}$ and $2_{on}$). Thus, both initial and final substates contribute to the rank $R_{z,z'}$ for $z \neq z'$, and the expressions for the $\varphi_{z,ij}(t)$ s are distinct in each of the following three regimes:

(*a*) For $n_x \notin \{O_x\}$ and $n_y \notin \{O_y\}$,

$$\varphi_{x,j_y i_x}(t) = \frac{1}{N_{x,n_x}} \sum_{m_y} P_{y,m_y}(ss)(\mathbf{V_y})_{n_x m_y} \tilde{f}_{m_x n_x}(t) \sum_{n_y \notin \{O_y\}} \omega_{n_y m_x}, \tag{B7}$$

and,

$$\varphi_{y,i_x j_y}(t) = \frac{1}{N_{y \in O_y, m_x}} \sum_{n_y} P_{x,m_x}(ss)(\mathbf{V_x})_{n_y m_x} f_{n_x n_y}(t), \tag{B8}$$

where $N_{y \in O_y, m_x} = \sum_{n_y \in \{O_y\}} P_{x,m_x}(ss)(\mathbf{V_x})_{n_y m_x}$, and we associate $n_x \to i_x$ and $m_x \to j_y$.

(b) For $n_x \notin \{O_x\}$ and $n_y \in \{O_y\}$,

$$\varphi_{x,j_y i_x}(t) = \frac{1}{N_{x,n_x}} \sum_{m_y} P_{y,m_y}(ss)(\mathbf{V_y})_{n_x m_y} f_{n_y n_x}(t), \tag{B9}$$

and,

$$\varphi_{y,i_x j_y}(t) = \frac{1}{N_{y,n_y}} \sum_{m_x} P_{x,m_x}(ss)(\mathbf{V_x})_{n_y m_x} f_{n_x n_y}(t), \tag{B10}$$

where we associate $n_y \to j_y$ and $n_x \to i_x$.

(c) For $n_x \in \{O_x\}$ and $n_y \in \{O_y\}$,

$$\varphi_{y,i_x j_y}(t) = \frac{1}{N_{y,n_y}} \sum_{m_x} P_{x,m_x}(ss)(\mathbf{V_x})_{n_y m_x} \tilde{f}_{m_y n_y}(t) \sum_{n_x \in \{O_x\}} \omega_{n_x m_y}, \tag{B11}$$

and,

$$\varphi_{x,j_y i_x}(t) = \frac{1}{N_{x \in O_x, m_y}} \sum_{n_x \in \{O_x\}} P_{y,m_y}(ss)(\mathbf{V_y})_{n_x m_y} f_{n_y n_x}(t), \tag{B12}$$

where we associate $n_y \to j_y$ and $m_y \to i_x$.

Now, we use $O_y$ and $O_x$ for expressing $R_{x,y}$. When $M_x < N_y$ and $\{O_x\}$ and $\{O_y\}$ are as defined above,

$$R_{x,y} = M_x - (O_x - O_y). \tag{B13}$$

This result can be generalized to the case of *J* groups in the underlying KS that are connected in the way defined above for the case of a single pair of groups. The generalized result reads,

$$R_{x,y} = M_x - \sum_j (O_{x,j} - O_{y,j}). \tag{B14}$$

These expressions imply that $\tilde{M}_x$ and $\tilde{N}_y$ on Eq. (10) are related to the KS's topology by,

$$\tilde{M}_x = M_x - \sum_j O_{x,j}, \tag{B15}$$

and,

$$\tilde{N}_y = \sum_j O_{y,j}. \tag{B16}$$

When $M_x > N_y$, and there are groups $\{Z_x\}$ and $\{Z_y\}$, with $Z_x < Z_y$, such that substates in $\{Z_y\}$ are connected only to substates in $\{Z_x\}$, we define $O_x = M_x - Z_x$ and $O_y = N_y - Z_y$, and Eq. (B13) holds. For *J* such groups, we define $O_{x,j} = M_x/J - Z_{x,j}$ and $O_{y,j} = N_y/J - Z_{y,j}$, and Eqs. (B14)-(B16) hold.

For a KS with symmetry, $\tilde{M}_x$ and $\tilde{N}_y$ are chosen in a different way than the one relies on the *on-off* connectivity; for such a case, the choice that makes the number of additives in the external sums of Eq. (10) minimal simply groups the identical PDFs. The topology of the RD form is determined by the largest $R_{x,y}$.

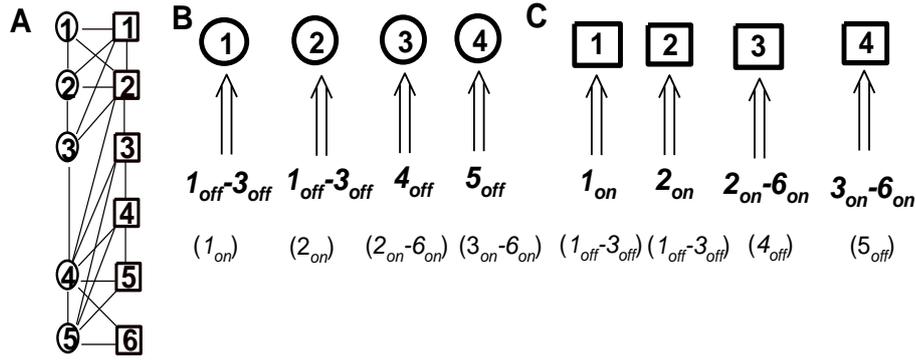

**FIG B4 A** A reversible connection KS with $R_{x,y}=4$ ($x \neq y$). The RD form's topology is shown on **B-C**. The clustering procedure and the parent substates (in the parenthesis) are indicated at the base of the double arrows. For example, substate $1_{off}$ in the RD form corresponds to the cluster of initial-*off*-substates $1_{off}$-$3_{off}$ in the KS. These are connected to substate $1_{on}$ in the KS. The WT-PDFs for the connections in the RD form can be obtained from Eqs. (B7)-(B12).

**Figure captions**

**FIG 1** A two-state trajectory (**A**), the KS (**B**), and the RD form (**C**). In trajectory **A**, a period in black is actually the occurrence of several very fast *on-off* cycles. The KS **B** has $L_{on}=2$ (squared substates), $L_{off}=2$ (circled substates), irreversible transitions, and $N_{on}=M_{on}=2$ and $N_{off}=M_{off}=2$. For generating the data, we take the following transition rate values ($k_{ji}$ connects substates $i \to j$): $k_{1_{off}1_{on}} = 0.3$, $k_{2_{off}2_{on}} = 0.02$, $k_{1_{on}1_{off}} = 0.425$, $k_{2_{on}1_{off}} = 0.075$, $k_{1_{on}2_{off}} = 0.0085$, $k_{2_{on}2_{off}} = 0.0015$ (with arbitrarily units). The equality of the ratios, $k_{2_{on}2_{off}}/k_{1_{on}2_{off}} = k_{2_{on}1_{off}}/k_{1_{on}1_{off}} \equiv p_R/p_L$ ($p_L + p_R = 1$), means that the KS is symmetric in the sense that the ranks of the 2D WT-PDFs of successive *x, y* (=*on,*

*off*) events are all equal one, except $R_{on,off}$ which equal two. As a result, the RD form (**C**) has one *on* substate and two *off* substates. The RD form has also direction dependent WT-PDFs for the *on* to *off* connections, $\varphi_{on,11}(t) = p_L k_{1_{off}1_{on}} e^{-k_{1_{off}1_{on}}t}$, $\varphi_{on,21}(t) = p_R k_{2_{off}1_{on}} e^{-k_{2_{off}1_{on}}t}$, and $\varphi_{off,11} = k_{1_{off}} e^{-k_{1_{off}}t}$, $\varphi_{off,12}(t) = k_{2_{off}} e^{-k_{2_{off}}t}$. Here, $k_i = \sum_j k_{ji}$, and $\varphi_{x,ij}(t)$ is the WT-PDF that connects substates $x_j \rightarrow y_i$ in the RD form.

**FIG 2** Indistinguishable KSs. KSs **A**-**C** have the simplest RD form (**D**) of one substate in each of the states. KSs **A**-**C** are equivalent when they have the same $\phi_{on}(t)$ and $\phi_{off}(t)$. Equivalent KSs **E**-**F** have $R_{x,y}$=2, *x, y = on, off*, and tri-exponential $\phi_{on}(t)$ and $\phi_{off}(t)$. The corresponding RD form is shown in **G**.

**FIG 3** Distinguishable KSs with $R_{x,y}$=2, *x, y = on, off* and bi-exponential $\phi_{on}(t)$ and $\phi_{off}(t)$. (We exclude symmetry in this example). KS **3C** is distinct from KSs **3A** and **3B**, because the corresponding RD forms, **3E** and **3D**, respectively, have different connectivity. KS **3A** and KS **3B** are also distinct, because the WT-PDFs for the connections in the RD form of KS **3A** are exponential, whereas those of KS **3B** are direction-dependent and bi-exponentials.

**FIG 4** The WT-PDFs of the *on* (left) and the *off* (right) states on a log-log scale. Shown are both the WT-PDFs obtained from a numerical solution of Eq. (2), full line, and by constructing these PDFs from a $10^6$ *on-off* event trajectory, circled symbols.

**FIG 5** The successive singular value ratio method for estimating the rank of two dimensional histograms. **A** - The successive singular value ratio method is applied on $\phi_{on,off}(t_1,t_2)$ and its first three cumulative PDFs. The first ratio contains most of the signal in all PDFs. **B –** The second and third order cumulative PDFs indicate a rank 2 PDF. **C –** The same method is applied on $\phi_{off,on}(t_1,t_2)$, and indicates a rank one PDF. Here the diamond and square symbols correspond to the second and first cumulative PDFs respectively, and the circled symbols correspond to the original PDF.

# Figures

**Figure 1**

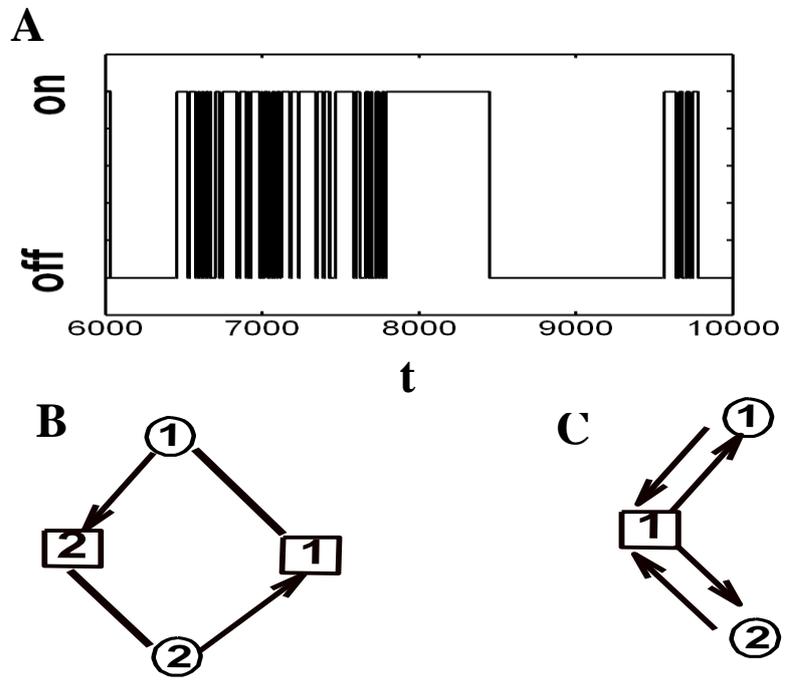

**Figure 2**

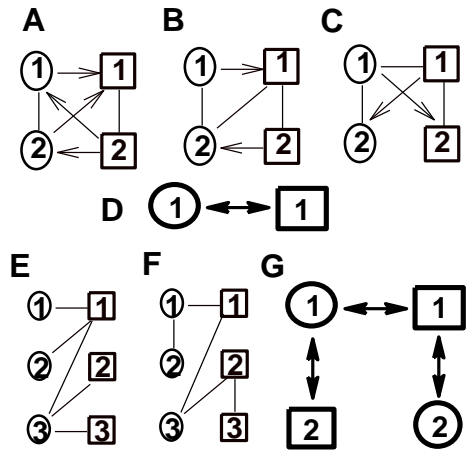

**Figure 3**

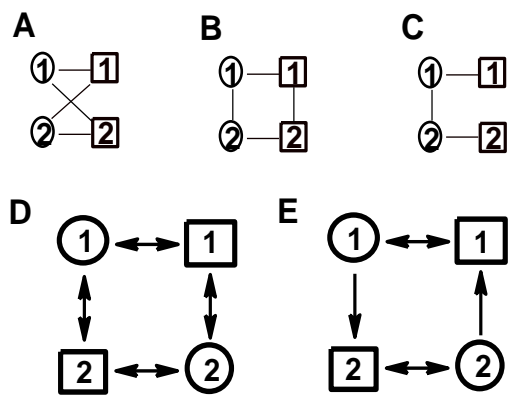

**Figure 4**

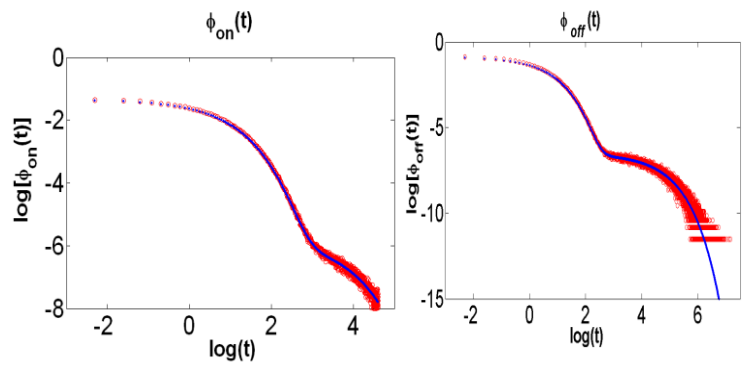

**Figure 5**

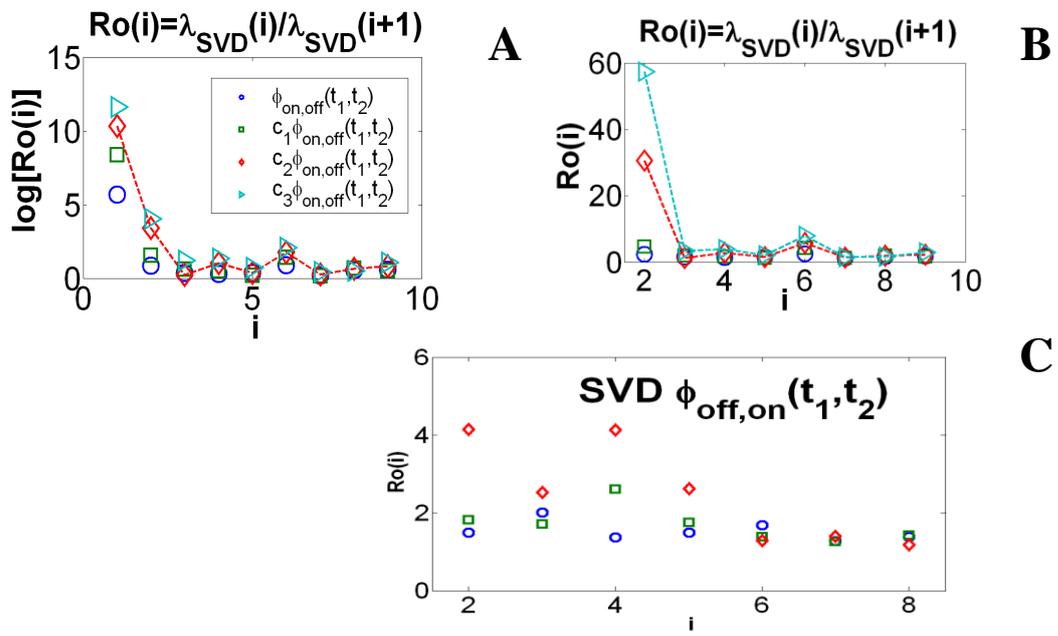